\documentclass[12pt]{article}
\usepackage{epic}
\newcommand{\be}{\begin{equation}}
\newcommand{\ee}{\end{equation}}
\newcommand{\bea}{\begin{eqnarray}}
\newcommand{\eea}{\end{eqnarray}}

\newcommand{\Li}[2]{{\mbox{Li}}_{#1}\left(#2\right)}
\def\cAdiv{  {\cal A}_{\rm div}  }
\def\cAdivSYM{ {\cal A}_{{\rm div,SYM}} }
\def\NP{{\rm NP}} 
\def\P{{\rm P}} 
\def\f{h}

\def\e{{\rm e}}

\def\eg{{\it e.g.}}
\newcommand{\Z}{\mathsf{Z}\kern -5pt \mathsf{Z}}
\def\half{ {1\over 2} }

\def\cO{  {\cal O}  }
\def\cN{  {\cal N}  }
\def\cM{  {\cal M}  }
\def\cI{  {\cal I}  }

\def\theequation{\thesection.\arabic{equation}}

\begin{document}

\begin{flushright}
BRX-TH-595\\
BOW-PH-142\\
TIT/HEP-582
\end{flushright}
\vspace{30mm}

\vspace*{.3in}

\begin{center}
{\Large\bf\sf  Two-loop graviton scattering relation \\
and IR behavior in $\cN=8$ supergravity }

\vskip 5mm Stephen G. Naculich\footnote{Research supported in part by the 
NSF under grant PHY-0456944}$^{,a}$,
Horatiu Nastase,$^{b}$ 
and Howard J. Schnitzer\footnote{Research supported in part 
by the DOE under grant DE--FG02--92ER40706\\
{\tt \phantom{aaa} 
naculich@bowdoin.edu, nastase@phys.titech.ac.jp, schnitzr@brandeis.edu}
}$^{,c}$
\end{center}

\begin{center}
$^{a}${\em Department of Physics\\
Bowdoin College, Brunswick, ME 04011, USA}

\vspace{.2in}

$^{b}${\em Global Edge Institute\\
Tokyo Institute of Technology, Tokyo 152-8550, Japan} 

\vspace{.2in}

$^{c}${\em Theoretical Physics Group\\
Martin Fisher School of Physics\\
Brandeis University, Waltham, MA 02454, USA}
\end{center}
\vskip 2mm

\begin{abstract}
We derive an ABDK-like relation 
between the one- and two-loop four-graviton amplitudes
in $\cN=8$ supergravity.
Specifically we show that the infrared divergent part of the
two-loop amplitude is one-half the square of the one-loop amplitude,
suggesting an exponential structure for IR divergences.
The difference between the two-loop amplitude and one-half the 
square of the full one-loop amplitude is therefore finite, and
expressible in a relatively simple form.
We give arguments for generalizations 
to higher loops and $n$-point functions, 
suggesting that the exponential of the 
full one-loop amplitude may be corrected, 
to low orders, by only simple finite terms. 

\end{abstract}

\vfil\break


\section{Introduction}
\setcounter{equation}{0}
\label{secintro}

Many advances have been made recently
in understanding the structure of the loop expansions of
${\cN}=4 $ super Yang-Mills 
(in particular, in completely computing its gluon scattering amplitudes)
and $\cN=8$ supergravity scattering amplitudes. 

The realization that the loop expansion of $\cN=4$ SYM amplitudes 
has an iterative structure began with the result of 
Anastasiou, Bern, Dixon, and Kosower (ABDK) 
relating the two-loop planar four-point gluon scattering amplitude 
to the one-loop amplitude \cite{Anastasiou:2003kj}
\be
\label{abdk}
M_4^{(2)} (\epsilon)  
=  \half \left[ M_4^{(1)} (\epsilon)  \right]^2 
  + \left( - \zeta_2 - \zeta_3 \epsilon + \cdots \right) 
 M_4^{(1)} (2 \epsilon)  
 + \hbox{const} 
+ \cO(\epsilon) \,.
\ee
Although $\cN=4$ SYM theories are UV finite,   
scattering amplitudes contain infrared divergences,
which are controlled by dimensional regularization
in $D= 4 - 2 \epsilon$ dimensions.
Subsequently, Bern, Dixon, and Smirnov (BDS) \cite{Bern:2005iz},
building on the work of refs.~\cite{Magnea:1990zb, Sterman:2002qn},
realized that the IR divergent factors of planar $n$-point amplitudes 
in $\cN=4$ SYM have an exponential form,
and are completely governed by two functions 
of the coupling $\lambda_{SYM}$:
the cusp anomalous dimension $f(\lambda_{SYM})$ 
and the collinear anomalous dimension $g(\lambda_{SYM})$.
They also conjectured a complete nonperturbative exponential ansatz 
for planar, MHV 
$n$-point scattering amplitudes in ref.~\cite{Bern:2005iz}.
In their ansatz for the four-point function,
the finite part is completely determined by the cusp 
anomalous  dimension $f(\lambda_{SYM})$.
The form of their full four-point ansatz was subsequently
confirmed in the large coupling limit 
using the AdS-CFT correspondence \cite{Alday:2007hr}. 
The nonperturbative form of $f(\lambda_{SYM})$ was computed 
in ref.~\cite{Beisert:2006ez}.
The BDS ansatz for the four and five-point function was also proved
using dual conformal symmetry \cite{Drummond:2007cf,Drummond:2007au}, 
while it was found that for six-point functions and above,
there are finite corrections to the BDS 
ansatz \cite{Alday:2007he,Drummond:2007bm,Bern:2008ap,Drummond:2008aq,Astefanesei:2007bk}. 
The form of the IR divergent factor for any $n$ was confirmed 
in ref.~\cite{Buchbinder:2007hm}.

In a parallel development, ${\cN}=8$ supergravity amplitudes
have been found to be much better behaved in the UV than previously thought, 
and generally to be
much simpler than a field theory of quantum gravity is a priori expected
to be.  
In particular, an explicit three-loop four-graviton 
scattering calculation found no UV divergences \cite{Bern:2007hh} 
and various arguments have been given that 
${\cN}=8$ supergravity is UV finite in four dimensions
up to eight loops \cite{Green:2006yu}
or even to all orders in perturbation theory
\cite{Bern:2006kd,Green:2006gt,Green:2006yu,Bern:2007hh}.
There may be, however, nonperturbative obstructions to UV finiteness
\cite{Green:2007zzb}.

Despite the fact that $\cN=4$ SYM theory 
is a (finite, superconformal) gauge theory 
and $\cN=8$ supergravity
a (potentially non-renormalizable) theory of quantum gravity,
there are deep connections between their 
perturbative scattering amplitudes.
Their tree-level amplitudes are closely related 
by the string theory relations of 
Kawai, Lewellen, and Tye (KLT) \cite{Kawai:1985xq}.
These tree-level relations were employed 
for loop calculations of ${\cN}=8$ supergravity amplitudes 
using unitarity methods \cite{Bern:1998ug}.

In this paper, we begin addressing the question: 
is it possible, in virtue of the  KLT relations, 
that the exponential structure of both the 
infrared divergent and the finite parts of the $\cN=4$ SYM amplitudes 
extends to ${\cN}=8$ supergravity amplitudes
at least up to the order to which ${\cN}=8$ supergravity is 
UV finite?\footnote{Since we work
in dimensional regularization, the UV and IR divergences will mix up
afterwards, making any structure harder to disentangle.}
In fact, we prove a relation 
for the four-graviton scattering amplitude
analogous to the ABDK relation (\ref{abdk})
\be
\label{ourresult}
M_4^{(2)} (\epsilon)  
=  \half \left[ M_4^{(1)} (\epsilon)  \right]^2 
+ {\rm finite}  + \cO(\epsilon) 
\ee
where the explicit form of the finite part 
is specified in eq.~(\ref{finalresult}).\footnote{After the
work described in this paper was completed, one
of the authors learned from Lance Dixon that he 
was previously aware of the relation (\ref{ourresult}).}
The relation (\ref{ourresult}) for $\cN=8$ supergravity is 
not as strong as the ABDK relation (\ref{abdk}), in which 
the finite part is actually a constant rather than a function
of the kinematic variables.

We make several observations about this result.
First, whereas the ABDK result (\ref{abdk}) only holds
in the large-$N$ limit, and therefore only involves
planar diagrams, the analogous result (\ref{ourresult}) for 
supergravity requires collusion between 
planar and non-planar diagrams.

Second, eq.~(\ref{ourresult}) implies that 
the IR-divergent part of the scattering amplitude through two loops
is given exactly by the exponential of the one-loop amplitude
(and as a result depends not only on the divergent but
also on the finite part of the one-loop amplitude).
This relation is actually simpler than that for $\cN=4$ SYM, 
where the two-loop divergences are modified by terms
proportional to the $\cO(\lambda_{SYM}^2)$ coefficients 
of $f(\lambda_{SYM})$ and $g(\lambda_{SYM})$.
The absence of such corrections in ${\cN}=8$ supergravity 
may be explained by the dimensionality of the gravitational 
coupling $\kappa$, which dictates that a term like
$M_4^{(1)} (2 \epsilon) $ would need to be multiplied 
by a function of $s$, $t$, and $u$ of degree one.
Cyclic symmetry, however, allows only $s+t+u$, which vanishes 
for massless gravitons.

Third, the finite remainder in  eq.~(\ref{ourresult}),
while apparently not expressible in terms of the one-loop amplitude,
is much simpler than the complete finite piece of the
two-loop amplitude itself, as we will see in section 2.
Hence, 
a large part of the finite two-loop amplitude 
is determined by the square of the one-loop amplitude.
It is therefore probably similar to the case of the six-point 
gluon amplitude in ${\cal N}=4$ SYM \cite{Bern:2008ap}.

The paper is organized as follows: 
in section 2, we perform the main calculation of this paper, 
obtaining the  ABDK-like relation (\ref{ourresult}) for 
the two-loop amplitude. 
In section 3,  we analyze more generally the IR behavior of 
the four-graviton amplitude,
and make some conjectures for higher $n$-point functions,
as well as for higher loop contributions.
Section 4 contains our conclusions. 

\section{Two-loop relation for the four-graviton amplitude}  
\setcounter{equation}{0}
\label{sectwoloop}
\def\text{}

The full all-loop-orders graviton four-point amplitude 
of $\cN=8$ supergravity 
is proportional to the tree-level four-point amplitude  \cite{Bern:1998ug}
\be
{\cM}_4 
=  {\cM}_4^{\rm tree} \left[ 1 + M_4^{(1)} + M_4^{(2)} + \cdots \right]
\ee
where $ {\cM}_4^{\rm tree} $ contains all the helicity 
information of the external gravitons, 
and $M_4^{(L)}$ is a scalar (momentum-dependent) factor 
appearing at $L$ loops.
In this section, we will prove the ABDK-like relation (\ref{ourresult})
between $M_4^{(1)}$ and $M_4^{(2)}$,
suggestive of  an exponential form for the full amplitude.

Due to the KLT relations \cite{Kawai:1985xq}, 
${\cN}=8$ supergravity graviton amplitudes 
are closely related to ${\cN}=4$ SYM gluon amplitudes,
and so loop amplitudes in ${\cN}=8$ supergravity 
can be expressed in terms of the same scalar integrals
that appear in ${\cN}=4$ SYM theory.
The one-loop four-graviton amplitude is given by \cite{Bern:1998ug}
\bea
\label{oneloop}
M_4^{(1)}  &=& 
-i 
\left( \kappa \over 2 \right)^2
stu \,
 \Bigl[  \cI_4^{(1)}(s,t)
       + \cI_4^{(1)}(s,u)  
       + \cI_4^{(1)}(t,u)  \Bigr] 
\eea
where $s=(k_1+k_2)^2$, $t=(k_1+k_4)^2$,  and $u=(k_1+k_3)^2$
are the usual Mandelstam variables, 
obeying $s+t+u=0$ for massless external gravitons,
and $\cI_4^{(1)}(s,t)$ corresponds to the scalar box integral
\begin{equation} 
\label{scalaronebox}
\cI_4^{(1)}(s,t)
= \cI_4^{(1)}(t,s)
= \mu^{4 - D}  \int {d^D p \over (2\pi)^D
} \;
{1\over p^2 (p-k_1)^2 (p-k_1-k_2)^2 (p+k_4)^2 } \, .
\nonumber\\
\end{equation}
We regularize loop integrals by evaluating them in 
$D=4-2 \epsilon$ dimensions.
In the region where $s$, $t<0$, 
the scalar box integral (\ref{scalaronebox}) is given by \cite{Bern:2005iz} 
\bea
\label{oneboxu}
\cI_4^{(1)}(s,t)
&=& 
\frac{ i \mu^{2 \epsilon} \e^{-\epsilon \gamma} (4\pi)^{-D/2} }
      {(-s)^{1+\epsilon}(-t)}
\Bigg\{ 
 \frac{4}{\epsilon^2} +\frac{2 L}{\epsilon} - \frac{4\pi^2   }{3} 
+ {\epsilon} 
\bigg( 
 2 \,\Li{3}{x}  
 + 2\,L\,\Li{2}{x}  
\\ && \hspace*{30mm}
- U L^2- \frac{1}{3} L^3 
- \pi^2 U - \frac{7\pi^2}{6} L 
-  \frac{34}{3} \zeta_3
\bigg)
+ \cO(\epsilon^2) \Bigg\}
\nonumber
\eea
where $x=-t/s$, $L= -\log(-x)= \log(s/t)$, and $U = \log(1-x)= \log(-u/s)$,
and we have explicitly written 
$\cO(\epsilon)$ terms that will be needed later. 
For now we drop the $\cO(\epsilon)$ terms to write
(again for $s$, $t<0$)
\bea
\label{snegtneg}
\cI_4^{(1)}(s,t)
&=&
\frac{ i \e^{-\epsilon \gamma} (4\pi)^{-D/2} } { s\,t}
\Bigg\{ 
 \frac{2}{\epsilon^2}\left(\frac{\mu^2}{-s}\right)^\epsilon
+\frac{2}{\epsilon^2}\left(\frac{\mu^2}{-t}\right)^\epsilon
- \log^2 \left( s \over t\right) -\frac{4\pi^2}{3}
+ \cO(\epsilon) 
\Bigg\}
\\
&=& \frac{ i \e^{-\epsilon \gamma} (4\pi)^{-D/2} } { s\,t}
\Bigg\{ 
 \frac{4}{\epsilon^2}
-\frac{2}{\epsilon} \log\left(-s \over \mu^2\right)
-\frac{2}{\epsilon} \log\left(-t \over \mu^2\right)
+ 2 \log\left(-s \over \mu^2\right) \log\left(-t \over \mu^2\right)
-\frac{4\pi^2}{3}
\Bigg\}\,.
\nonumber
\eea
If we wish to evaluate this in the region $t>0$ and $s<0$, we
continue $t$ from the negative to the positive real axis 
in the upper half plane to obtain
\bea
\label{snegtpos}
&&\cI_4^{(1)}(s,t)
=
\frac{ i \e^{-\epsilon \gamma} (4\pi)^{-D/2} } {s\,t}
\Bigg\{ 
 \frac{4}{\epsilon^2}
-\frac{2}{\epsilon} \log\left(-s \over \mu^2\right)
-\frac{2}{\epsilon} 
\bigg[ \log \left( t \over \mu^2\right) - i \pi \bigg]
\nonumber\\&&\hspace{47mm}
+ 2 \log\left(-s \over \mu^2\right) 
\bigg[ \log\left(t \over \mu^2\right) - i \pi \bigg]
-\frac{4\pi^2}{3}
+ \cO(\epsilon) 
\Bigg\}\,.
\eea
Therefore, 
using the expression (\ref{snegtneg}) or (\ref{snegtpos})
as appropriate for each term,
we write the full one-loop scattering amplitude 
(\ref{oneloop}) in the physical region $t>0$ and $s$, $u<0$, 
\bea
\label{oneloopresult}
&& \hspace{-22mm}
M_4^{(1)} 
= \frac{\lambda}{8 \pi^2}
\Bigg\{
\frac{1}{\epsilon}
\bigg(
s\log\left(-s \over \mu^2\right)
+t \bigg[ \log\left(t\over \mu^2\right) - i \pi \bigg]
+u\log\left(-u\over \mu^2\right)
\bigg)
\nonumber\\&& 
+s \log \left(-u \over \mu^2\right)
\bigg[ \log \left(t \over \mu^2\right) - i \pi \bigg]
+t \log \left(-u \over \mu^2\right) \log \left(-s \over \mu^2\right)
\nonumber\\&&
+u \log \left(-s \over \mu^2\right) 
\bigg[ \log \left(t \over \mu^2\right) - i \pi  \bigg]
+ \cO(\epsilon) 
\Bigg\}
\eea
where
\be
\lambda=\left( \kappa\over 2 \right)^2
\left( 4\pi \e^{-\gamma} \right)^\epsilon \,.
\ee
The one-loop scattering amplitude in the region $s>0$ and $t$, $u<0$ 
may be obtained by simply exchanging $s \leftrightarrow t$ in
eq.~(\ref{oneloopresult}).
Note that, despite the $1/\epsilon^2$ divergence
 of the scalar loop integral $\cI_4^{(1)}(s,t)$,
the full one-loop four-graviton amplitude only has a
$1/\epsilon$ IR divergence.
This is as expected for gravity, as discussed in sec.~3 of this paper.

The one-loop expression (\ref{oneloopresult}) may be written in 
a completely permutation symmetric way as
\bea
\label{oneloopresulte}
&& \hspace{-12mm}
M_4^{(1)} 
=
\frac{\lambda}{8 \pi^2}
\Bigg\{
\frac{1}{\epsilon}
\left[
s\log\left(-s \over \mu^2\right)
+t\log\left(-t\over \mu^2\right)
+u\log\left(-u\over \mu^2\right)
\right]
\\ && 
+\,s \log \left(-t \over \mu^2\right) \log \left(-u \over \mu^2\right)
+\,t \, \log \left(-u \over \mu^2\right) \log \left(-s \over \mu^2\right)
+\,u \log \left(-s \over \mu^2\right) \log \left(-t \over \mu^2\right)
+ \cO(\epsilon) 
\Bigg\}
\nonumber
\eea
an expression which is manifestly real in the Euclidean region $s,t,u<0$.

Now we turn to the two-loop four-point graviton amplitude \cite{Bern:1998ug}
\be
\label{twoloopamp}
M_4^{(2)}= 
\left( \kappa \over 2 \right)^4 
s^3tu \, 
\Bigl[\cI_4^{(2) \P}(s, t)
+ \cI_4^{(2) \P}(s, u)  
+ \cI_4^{(2) \NP}(s, t)
+ \cI_4^{(2) \NP}(s, u)
\Bigr] 
+ \left({\hbox{cyclic perms}\atop \hbox{of~} s,t,u} \right)
\ee
which receives contributions both from the scalar double-box integral
\begin{equation}
 \cI_4^{(2)  \P}(s,t) = 
\mu^{8 - 2 D}
\int {d^{D}p\over (2\pi)^{D}} \; {d^{D}q\over (2\pi)^{D}} \;
 {1\over p^2 \, (p + q)^2 q^2 \, (p - k_1)^2 \,(p - k_1 - k_2)^2 \,
        (q-k_4)^2 \, (q - k_3 - k_4)^2 } 
\end{equation}
as well as from the two-loop non-planar integral
\begin{equation} 
\cI_4^{(2)  \NP}(s,t)  = 
\mu^{8 - 2 D}
\int {d^{D} p \over (2\pi)^{D}} \, {d^{D} q \over (2\pi)^{D}} \
{1\over p^2\,(p+q)^2\, q^2 \, (p-k_2)^2  \,(p+q+k_1)^2\,
  (q-k_3)^2 \, (q-k_3-k_4)^2}  \,.
\end{equation} 
The non-planar integral has been evaluated by Tausk \cite{Tausk:1999vh}, 
who writes it as
\begin{equation}
\label{tausk}
\cI_4^{(2)  \NP}(s,t)
= \cI_4^{(2)  \NP}(s,u)
= - (4\pi)^{-D} \Gamma(1+\epsilon)^2 
 \left\{ \frac{F_t}{s^2 t} + ( t \leftrightarrow u ) \right\}
\end{equation}
where the expression for $F_t$ takes different forms in 
different regions depending on the signs of $s$, $t$, and $u$. 
We use eq.~(\ref{tausk}) to re-express 
the two-loop amplitude (\ref{twoloopamp})  as
\begin{equation}
\label{twoloopampall}
M_4^{(2)}=
\left( \kappa \over 2 \right)^4 
s^3tu \, 
\left[\cI_4^{(2) \P}(s, t)
- 2 (4\pi)^{-D}  \Gamma(1+\epsilon)^2
 \frac{F_t}{s^2 t} 
\right] 
+ \left({\hbox{all perms}\atop \hbox{of~} s,t,u} \right)\,.
\end{equation}
We begin with an expression \cite{Smirnov:1999gc, Bern:2005iz} 
for the scalar double-box integral 
in the region $s$, $t<0$ (hence $u >0$):
\bea
\label{twoboxu}
\cI_4^{(2) \P}(s,t)
&=&
\left( i \mu^{2\epsilon} \e^{-\epsilon \gamma} (4\pi)^{-D/2}  \right)^2
\frac{1}{(-s)^{2+2\epsilon}(-t)}
\Bigg\{ 
-\frac{4}{\epsilon^4} -\frac{5 L}{\epsilon^3}
+ \frac{1}{\epsilon^2} \bigg( - 2 L^2 +\frac{5\pi^2}{2} \bigg) 
\nonumber \\ && 
+ \frac{1}{\epsilon} \bigg( -4 \, \Li{3}{ x } - 4\,   L \, \Li{2}{ x }
+ 2\, U  L^2  +\frac{2}{3} L^3 +2 \pi^2 U +\frac{11\pi^2}{2} L
+\frac{65}{3} \zeta_3 \bigg) 
\nonumber \\
&& 
+ \, 44 \, \Li{4}{ x }  
-4 \, S_{2,2}(x)  
+ \biggl(24\,L -4\, U   \biggr) \Li{3}{ x }
-4  \, L\, S_{1,2}(x)  
\nonumber \\ && 
+ \bigg( 2\, L^2 -4 \, U  L  +\frac{20\pi^2}{3}\bigg) \Li{2}{x}
+U^2 L^2 +\frac{8}{3} U L^3 +\frac{4}{3}  L^4  
\nonumber \\ &&
+\biggl( U^2 + \frac{10}{3} U L +6 L^2  \biggr) \pi^2 
+ \biggl(4\,U +\frac{88}{3}\,L \biggr) \zeta_3  +\frac{29\pi^4 }{30}
+ \cO(\epsilon) \Bigg\}\,.
\eea
It will be convenient to evaluate $M_4^{(2)}$ 
in a region where $t>0$ and $s$, $u<0$,
and therefore we must analytically continue eq.~(\ref{twoboxu})
into this region. 
To do so, first we re-express the generalized polylogarithms 
$\Li{n}{x}$ and  $S_{n,p}(x)$ appearing in eq.~(\ref{twoboxu})
as functions of $y \equiv 1/x$ using identities (\ref{appone})
given in the appendix.
Next, we analytically continue $t$ from the negative to the
positive real axis through the upper half plane (holding $s$ fixed),
which takes $L \to -T + \pi i$ and $U \to V + T - \pi i$, 
where 
$T= \log(x)= -\log(y) = - \log(-s/t)$ and  
$V = \log(1- y) = \log(-u/t)$.
After the continuation, we have $0<y< 1$, 
so that polylogarithms with argument $y$ 
do not pick up additional contributions from the analytic
continuation (since the branch cut for polylogarithms 
along the positive real axis starts to the right of unity).
Finally, we write 
$(-s)^{-2\epsilon} = t^{-2\epsilon} \exp(2 \epsilon T)$ to obtain 
\bea
\label{twoboxt}
\cI_4^{(2) \P}(s,t)
&=&
\frac{\mu^{4 \epsilon} \e^{-2 \epsilon \gamma} (4\pi)^{-D} }
{s^2 t^{1+2\epsilon}}
\Bigg\{  -\frac{4}{\epsilon^4} 
+\frac{1}{\epsilon^3} \left( -3\,T - 5 \pi i\right)
+ \frac{1}{\epsilon^2}\bigg(\frac{9\pi^2}{2} - 6\pi i T\bigg)
\nonumber\\ &&
+ \frac{1}{\epsilon} 
\bigg( -4 \,\Li{3}{y} -4\,T\, \Li{2}{y} 
+\frac{2}{3}  T^3 +2 V T^2 +\frac{11\pi^2}{2}  T +\frac{65 \zeta_3}{3}\bigg)
\nonumber\\ && 
+ \frac{i \pi}{\epsilon}
\bigg( 4 \, \Li{2}{y}  -2   \,T^2 -4  \, V T +\frac{7\pi^2 }{2}   \bigg)
-36 \,\Li{4}{y} -4 \,S_{2,2}(y)
\nonumber\\ && 
-4\,T\,  S_{1,2}(y)
+ \biggl(-28\,T - 4 \, V \biggr)  \Li{3}{y}
+\bigg(-10 T^2 - 4\, V T -\frac{14\pi^2}{3}  \bigg) \Li{2}{y}
\nonumber \\  &&
+2 V T^3 +V^2 T^2
+\bigg( \frac{14}{3}  V T +\frac{7}{3} T^2 \bigg) \pi^2
+\biggl( 18\, T + 4\, V \biggr) \zeta_3 
-\frac{113\pi^4}{90} 
\nonumber\\ &&
+i \pi \bigg[ 20 \,  \Li{3}{y} +4 \, S_{1,2}(y) +
\biggl(4 \,  V +12\,T \biggr)  \Li{2}{y} 
\nonumber \\ &&
+\frac{4}{3}   T^3 -2   V T^2 -2   V^2 T
+\left(\frac{19}{3} T -\frac{2}{3} V\right) \pi^2 
+\frac{76}{3}   \zeta_3
 \bigg]
+ \cO(\epsilon) \Bigg\}
\eea
valid in the region $t>0$ and $u$, $s<0$.
In the same region, the non-planar integral is given by \cite{Tausk:1999vh}
\bea
\label{tauskF}
F_t &=& 
\left(\mu^2 \over t\right)^{2\epsilon} \Bigg\{ \vphantom{\frac{1}{1}} 
- \frac{2}{\epsilon^4}
+ \frac{1}{\epsilon^3} 
\bigg( 2\,T + \frac{7}{2} V - \frac{5\pi i }{2} \bigg)
\nonumber\\ && 
+ \, \frac{1}{\epsilon^2}\,  \bigg(
2\,T^2+T V -V^2 +  6\,T+6\,V + \frac{31\pi^2}{12}
 + i \pi \left[ T + 4\,V \right]
\bigg)
\nonumber \\ && 
+\,  \frac{1}{\epsilon}\,  \bigg(
2\,  S_{1,2}(y)  
-\frac{2}{3} T^3 -2\,T^2 V
   -2\,T V^2 -V^3
  -24\,T-24\,V
 +\left(-\frac{23}{6}\,T-\frac{41}{6}\,V \right) \pi^2
\nonumber 
\\ && 
\hspace{8mm}
+ \frac{15}{2} \zeta_3 
+ i \pi \left[ - 2\,\Li{2}{y} 
+ 4\,T^2 + 2\, T V -3\, V^2
+\, 12\,T + 12\,V + \frac{5\pi^2}{2} \right]
\bigg)
\nonumber \\ &&
+ 12\,\Li{4}{y}  -62\,  S_{2,2}(y) + 26\,  S_{1,3}(y)
 + \biggl(12\,T-18\,V +24\biggr) \,\Li{3}{y}
\nonumber \\ &&
 + \biggl(- 44\,T + 6\,V + 24 \biggr) \,  S_{1,2}(y)
 +\biggl(6\,T^2-18\,T V+ 24\,T +13 \pi^2 \biggr) \, \Li{2}{y}
\nonumber \\ &&
 -\, \frac{11}{6} T^4 -\frac{13}{3} T^3 V + T^2 V^2 +2\,T V^3
 +\frac{4}{3} V^4
-4\,T^3 -12\,T^2 V-4\,V^3
\nonumber \\ &&
 + \left( -\frac{22}{3} T^2
 -\frac{23}{3} T V + \frac{37}{6} V^2 
-14\,T-14\,V \right) \pi^2
+96\,T + 96\,V 
 -\frac{311\pi^4}{120} 
\nonumber \\ &&
 -\biggl(45\,T+39\,V+24\biggr)\, \zeta_3
 +  i \pi \bigg[ 32\,\Li{3}{y} + 28\,  S_{1,2}(y) 
 + \biggl(  14\,T +12\, V - 48 \biggr) \Li{2}{y}
\nonumber \\ &&
 +\, \frac{11}{3} T^3 +2\,T^2 V + \frac{4}{3} V^3
+ 12\,T^2 +24\,T V -12\,V^2
 -48\,T -48\,V 
\nonumber \\ && 
+ (- 6V + 4 ) \pi^2 -15\,\zeta_3
\vphantom{\frac{1}{1}}
\bigg] 
+ \cO(\epsilon) \Bigg\}
\, .
\eea
Inserting eqs.~(\ref{twoboxt}) and (\ref{tauskF}) 
into eq.~(\ref{twoloopampall}), and using 
\begin{equation}
\Gamma(1+\epsilon)^2 
= \e^{-2 \epsilon \gamma} \left(
1 + {\pi^2 \over 6} \epsilon^2 - {2\over 3} \zeta_3 \epsilon^3
+ {7 \pi^4\over 360} \epsilon^4 + \cdots \right)
\end{equation}
we obtain 
\bea
\label{planarandnonplanar}
&&\hspace{-8mm}
M_4^{(2)}=
\frac{\lambda^2 su }{(4\pi)^4}
\left(\frac{\mu^2}{t}\right)^{2\epsilon}
\Bigg\{
 - \frac{7}{\epsilon^3} \left( T+V  \right)
+\frac{1}{\epsilon^2} \biggl( 
 2\, V^2-2\, V T -4\, T^2
-12\, T -12\, V -8 i \pi \left[ T +V \right]
\biggr)
\nonumber\\&& 
+ \frac{1}{\epsilon} \bigg(
-4 \,\Li{3}{y} -4 \, S_{1,2}(y) -4\,T\, \Li{2}{y}
+2 \, T^3 +6 \, V T^2 +4 \, V^2 T +2 \, V^3
\nonumber\\&& 
+\bigg( \frac{25\pi^2}{2}  +48 \bigg) (T+V)
+4 \zeta_3
+i\pi 
\bigg[ 
8  \,\Li{2}{y}
-10  T^2 -8  V T +6  V^2
-24  (T +  V)
-\frac{2\pi^2}{3}
\bigg]
\bigg)
\nonumber\\&& 
-60 \,\Li{4}{y}
+120 \, S_{2,2}(y)
-52 \, S_{1,3}(y)
+\biggl( -52 T + 32 V  -48 \biggr) \Li{3}{y}
\nonumber\\&&
+\biggl(84 \,T -12 V-48 \biggr) S_{1,2}(y)
+ \biggl( -22\, T^2 +32 V T -48 \, T 
-\frac{92\pi^2}{3} \biggr) \Li{2}{y}
+ \frac{11}{3}T^4 
\nonumber\\&&
+\frac{32}{3}V T^3
-V^2 T^2 
-4 V^3 T
-\frac{8}{3} V^4
+8 T^3
+24 V T^2
+8 V^3
+\bigg( \frac{49 }{3} T^2 +\frac{59}{3}  V T -12  V^2\bigg) \pi^2
\nonumber\\&& 
+( 26 \pi^2 - 192) (T+ V) 
+\bigg( \frac{332}{3} T +\frac{260}{3} V+ 48 \bigg) \zeta_3
+\frac{283 \pi ^4}{90}
+ i\pi \bigg[
-44  \,\Li{3}{y} -52   \, S_{1,2}(y)
\nonumber\\&& 
+\biggl( -20  V -16 \,T +96 \biggr) \Li{2}{y}
-6  T^3
-6  V T^2
-2  V^2 T
-\frac{8}{3}  V^3
-24  T^2 
-48  V T
+24  V^2 
\nonumber\\ && 
+( 6\, T +10 \,V ) \pi^2 
+96  (T+V)
+52   \zeta_3
-8 \pi^2
\bigg]
+ \cO(\epsilon) \Bigg\}
+ \left({\hbox{all perms}\atop \hbox{of~} s,t,u} \right)\,.
\eea
Note that the leading $1/\epsilon^4$ pole of the 
planar integral (\ref{twoboxu}),
which is present in the two-loop $\cN=4$  SYM amplitude,
is cancelled in the two-loop $\cN=8$ supergravity amplitude
by the $1/\epsilon^4$ pole of the two-loop non-planar integral.
Additional cancellations of poles will occur when we add the
other permutations of $s$, $t$, and $u$.

First we consider the permutation that exchanges $s$ and $u$, 
which can be obtained by simply letting
$y \to 1-y$, $T\to -V$,  and $V \to -T$ in the expression above 
(without any analytic continuation required --- this is 
the reason we chose to evaluate the amplitude in the region 
$t>0$ and $s$, $u<0$).
Adding eq.~(\ref{planarandnonplanar}) and the $(s\leftrightarrow u)$ 
permutation,  
and using identities (\ref{apptwo}) relating 
polylogarithms with argument $1-y$ to those with argument $y$,
we obtain
\bea
\label{simple}
&& \hspace{-12mm}
M_4^{(2)}
= \frac{\lambda^2 su}{(4\pi)^4}
\left(\frac{\mu^2} {t}\right)^{2\epsilon}
\Bigg\{
- \frac{2}{\epsilon^2} \left(  T+V \right)^2
-\frac{4 \pi i }{\epsilon} \left(  T+V \right)^2
-8 \,\Li{4}{y} +8 \, S_{1,3}(y)
\nonumber\\&&
+\left( -4 V -12 \,T \right) \, \Li{3}{y}
-4 \left( T +V \right) S_{1,2}(y)
+\left( -8\, T^2 -4\,  V  T \right) \Li{2}{y}
\nonumber\\&&
+ \, T^4 +4 \, V T^3 +2 \, V^2 T^2 +\frac{2 }{3}V^3 T +V^4
+\frac{13 \pi^2}{3} \left( T+V\right)^2 +4 (T+V) \zeta_3 +\frac{4 \pi ^4}{15}
\nonumber\\&&
+ i \pi \bigg[ 8 \,\Li{3}{y} -8 \, S_{1,2}(y)
+ \left( 16 \, T +8 \, V \right) \Li{2}{y}
-\frac{10}{3}  T^3 -2  V^2 T -10  V T^2 +\frac{10}{3}  V^3
\nonumber\\&& \hspace{8mm}
-\frac{2\pi^2 }{3} (T+V) +8  \zeta_3
\bigg]
+ \cO(\epsilon) \Bigg\}
+ \left({\hbox{cyclic perms}\atop \hbox{of~} s,t,u} \right) \,.
\eea
At this point, the $1/\epsilon^3$ pole has also cancelled, 
leaving an expression whose leading divergence is $1/\epsilon^2$.
This is as expected for a two-loop gravity amplitude, as discussed
in sec.~3.

Observe that one can define a degree of transcendentality 
for each term in an expression, with 
$\log^k z$, $\Li{k}{z}$, $S_{n,k-n}(z)$, $\zeta_k$, and $\pi^k$
(since $\zeta_{2m} \sim \pi^{2m}$) all having degree $k$,
where $z$ is any ratio of momentum invariants (\eg, $x$ or $y$). 
The degree of transcendentality is preserved by all 
(generalized) polylogarithm identities, and therefore well-defined.
Both the one-loop (\ref{oneloopresult}) and two-loop (\ref{simple})
results satisfy a simple rule: 
all terms proportional to $(\lambda/\epsilon^2)^L\cdot 
\epsilon^k$ have degree of transcendentality $k$, 
where $L$ is the loop order. 
Note, however, that while the one- and two-loop planar integrals 
(\ref{oneboxu}) and (\ref{twoboxu}) also satisfy this rule, 
the two-loop nonplanar integral (\ref{tauskF}) does not,
as it contains terms of subleading transcendentality.
The terms of subleading transcendentality only cancel out 
when we add the $u\leftrightarrow s$ permutation
in eq.~(\ref{simple}).\footnote{We would like to 
thank Lance Dixon for pointing out to us that this 
fact may not be widely known.}
We used this cancellation of terms of subleading transcendentality 
as a useful check on our intermediate calculations.
It remains an interesting question whether this 
``conservation law for transcendentality" 
persists to higher orders in the loop expansion.

Note also that the coefficients of 
the $1/\epsilon^2$ and $1/\epsilon$ poles of the amplitude
are considerably simpler than those of the 
original planar and non-planar integrals that contributed to it.
Its form suggests that it may be related to the square of the
one-loop amplitude (\ref{oneloopresult}), as we will now show.

Using $s+t+u=0$, we may express the  square of the one-loop amplitude 
(\ref{oneloop}) as 
\begin{equation} 
\label{square}
\left[M_4^{(1)}\right]^2=
\left( \kappa \over 2 \right)^4
su \left[ s t \, \cI_4^{(1)}(s,t) -u t\, \cI_4^{(1)}(u,t)   \right]^2
+ \left({\hbox{cyclic perms}\atop \hbox{of~} s,t,u} \right)\,.
\end{equation} 
Continuing the expression (\ref{oneboxu}) 
to the region $t>0$ and $s$, $u<0$ as we did before
for the two-loop amplitude, 
we find
\bea
\label{oneboxt}
\cI_4^{(1)}(s,t)
&=&
\frac{- i \mu^{2\epsilon} \e^{-\epsilon \gamma} (4\pi)^{-D/2} }
{(-s) t^{1+\epsilon}}
\Bigg\{  
 \frac{4}{\epsilon^2} 
+\frac{1}{\epsilon} \biggl( 2\,T + 2 \pi i\biggr)
+ \bigg(-\frac{4\pi^2}{3}  + 2\pi i T\bigg)
\nonumber\\ &&
+ {\epsilon} 
\bigg(  2 \,\Li{3}{y} +2\,T\, \Li{2}{y} 
 -\frac{1}{3}  T^3 - T^2 V - \frac{7\pi^2}{6} T -\frac{34 \zeta_3}{3}
\nonumber\\ && 
+ i \pi
\bigg[-  2 \, \Li{2}{y}  +  \,T^2  + 2 \, T V -\frac{\pi^2}{6} \bigg] 
\bigg)
+ \cO(\epsilon^2) \Bigg\}\,.
\eea
For $u<0$ and $t>0$,  we may find $\cI_4^{(1)}(u,t)$ by simply 
letting $y \to 1-y$,  $T\to -V$,  and $V\to -T$ in eq.~(\ref{oneboxt}). 
Using the identities (\ref{apptwo}), we then obtain
\bea
\label{oneloopdiff}
&&\hspace{-18mm}
s t \, \cI_4^{(1)}(s,t) -u t\, \cI_4^{(1)}(u,t)  
= \frac{ i \mu^{2\epsilon} \e^{-\epsilon \gamma} (4\pi)^{-D/2} }{ t^{\epsilon}}
\Bigg\{  
\frac{2}{\epsilon} (T+V) + 2\pi i (T + V )
\nonumber \\
&&  \hspace{5mm}
+\, {\epsilon}  \,
\bigg(  
 2 \, \Li{3}{y} +2 \, S_{1,2} (y) +2 \, T\, \Li{2}{y} 
-\frac{1}{3}T^3 - T^2 V -\frac{1}{3}V^3 -\frac{7 \pi ^2 }{6} (T+V)
\nonumber\\ 
&&  \hspace{5mm}
-2 \zeta_3
+ i \pi \left[ -4 \, \Li{2}{y} +  T^2 +2 \,  V T -  V^2 +\frac{\pi^2}{3}
\right]
\bigg)
+ \cO(\epsilon^2) \Bigg\}\,.
\eea
Inserting this result in eq.~(\ref{square}), we find that the
difference between the two-loop amplitude and half of
the square of the one-loop amplitude is {\it finite}, 
expressible in the rather compact form
\bea
\label{compact}
M_4^{(2)}-\frac{1}{2}\left[M_4^{(1)}\right]^2
&=&  \left( \kappa \over 8 \pi \right)^4 su
\Bigg\{
8 \, S_{1,3} (y) + \frac{1}{3} \log^4 y + 8 \zeta_4
+ i \pi \left[-8\, S_{1,2} (y) + \frac{4}{3} \log^3 y + 8 \zeta_3 \right]
\nonumber\\ && \hspace{20mm}
+ (y \to 1-y) 
\Bigg\}
+ \left({\hbox{cyclic perms}\atop \hbox{of~} s,t,u} \right)\,.
\eea
This relatively simple expression suggests that a large portion 
of the rather complicated finite piece of the two-loop amplitude 
(\ref{simple}) comes from the square of the one-loop amplitude, 
in particular involving nontrivially both the 
finite term and the $\cO(\epsilon)$ term.
The difference (\ref{compact}) may be rewritten as 
\bea 
\label{difference}
M_4^{(2)}-\frac{1}{2}\left[M_4^{(1)}\right]^2
&=&  \left( \kappa \over 8 \pi \right)^4 
\Bigg\{
  su \, \biggl[ \f(t,s,u) + \f(t,u,s) \biggr] 
+ tu \, \biggl[ \f(s,t,u) + \f(s,u,t) \biggr] 
\nonumber \\&& \hspace{15mm}
+ st \, \biggl[ \f(u,s,t) + \f(u,t,s) \biggr] 
\Bigg\}
\eea
where $\f(t,s,u)$ is given in the region $t>0$ and $s$, $u<0$ by the
expression
\be 
\label{ftsu}
\f(t,s,u) =
8 \, S_{1,3} (-s/t) + \frac{1}{3} \log^4 (-s/t) + 8 \zeta_4
+ i \pi \left[-8\, S_{1,2} (-s/t) + \frac{4}{3} \log^3 (-s/t) 
+ 8 \zeta_3 \right]\,.
\ee
To obtain an explicit expression for (\ref{difference}) 
we must analytically continue 
eq.~(\ref{ftsu}) into several other regions. 
To obtain $\f(s,t,u)$, we must first analytically continue
$\f(t,s,u)$ to the region where $s>0$ and $t$, $u<0$
(the explicit expression is in the appendix), 
and then exchange $s$ and $t$ in the resulting expression.
To obtain $\f(s,u,t)$, we must first analytically continue
$\f(t,s,u)$ to the region where $u>0$ and $s$, $t<0$ 
(also in the appendix), and then permute $s \to u \to t \to s$ in the result.
Using several additional identities, we may combine these pieces 
to obtain our final result
\bea
\label{finalresult}
&&\hspace{-10mm} 
 M_4^{(2)}-\frac{1}{2}\left[M_4^{(1)}\right]^2
\nonumber \\ 
&& = \left( \kappa \over 8 \pi \right)^4 
\Bigg\{
su
\bigg(
8 \, S_{1,3} (y) + \frac{1}{3} \log^4 y + 8 \zeta_4
+ i \pi \left[-8\, S_{1,2} (y) + \frac{4}{3} \log^3 y + 8 \zeta_3 \right]
\bigg)
\nonumber \\ && 
+  tu
\bigg(
  8\, S_{2,2} (y) 
- 8\, S_{1,3} (y) 
- 8\, \log y \, S_{1,2} (y)
- 4\pi^2 \, \Li{2}{y} 
+\frac{1}{3} \log^4 (1-y)
\nonumber \\  && \
-\frac{4}{3} \log y \, \log^3 (1-y)
+2 \log^2 y \, \log^2 (1-y)
+2 \pi^2  \log^2 (1-y)
-4 \pi^2 \log y \, \log (1-y)
\nonumber \\ && \
+i \pi \biggl[
\,8 \, S_{1,2}(1-y) +\frac{8\pi^2}{3} \log (1-y) - 8 \zeta_3 \biggr]
\bigg) 
\Bigg\}
+ \left( s \leftrightarrow u,  y \to 1-y \right)
\eea
valid in the physical region $t>0$ and $s$, $u<0$.
If we wish to obtain the result in the region
$s>0$ and  $t$, $u<0$, we simply exchange $s \leftrightarrow t$ 
which means that $y  = -s/t$ is replaced by $x = -t/s$ throughout
the expression above.

The function (\ref{ftsu}) can be written rather elegantly 
using eq.~(\ref{polyl}) as 
\bea
\label{ftsuint}
\hspace{-2cm}&&
\f(t,s,u) 
=16\zeta_4
\\  \hspace{-2cm}&&\hspace{1cm}
+\int_1^y \frac{ dy  }{y} \left[ 
\frac{4}{3} 
\left(\log^3(e^{i\pi}y)-\log^3(e^{i\pi}(1-y)) \right)
+4 \pi^2 
\left(\log(e^{i\pi}y)-\log(e^{i\pi}(1-y)) \right)
\right]\,.
\nonumber
\eea
The  expression (\ref{finalresult}) 
is not manifestly permutation symmetric in $s$, $t$, and $u$ 
since in the physical region in which we are working,
$t>0$ whereas $u, s<0$.
However, if we analytically continue\footnote{Note that 
the analytical continuation between various regions can be taken as 
$-s=se^{-i\pi}$ and similarly for $t$ and $u$, i.e. 
$-s=|s|e^{-i\pi \theta(s)}$, $s=|s|e^{+i\pi\theta(-s)}$ 
and similarly for $t$ and $u$.
We have checked that this gives the correct 
continuation of the planar and nonplanar integrals in 
refs.~\cite{Bern:2005iz,Tausk:1999vh}. }
this expression to the Euclidean domain $s$, $t$, and $u<0$,
then eq.~(\ref{finalresult}) becomes
\bea
&& \hspace{-15mm}
M_4^{(2)}-\frac{1}{2}\left(M_4^{(1)}\right)^2
= 4 \left( \kappa \over 8 \pi \right)^4 
su \Bigg\{4\zeta_4+\int_1^{s/t}
d\left[\log\left(\frac{s}{t}\right)\right]
\left[\frac{\log^3(\frac{s}{t})-\log^3(\frac{u}{t})}{3}
+\pi^2\log\frac{s}{u}\right]
\nonumber\\&& \hspace{40mm}
+(u\leftrightarrow s)\Bigg\}
+ \left({\hbox{cyclic perms}\atop \hbox{of~} s,t,u} \right) \,.
\eea
This expression is now explicitly 
symmetric in $s,t,u$ 
(since $s,t,u<0$ there is nothing to break the symmetry).
In order to go back to the polylogarithm form we must choose 
a Euclidean region constraint (thus breaking the symmetry).
Choosing $-|t|=u+s$, we obtain 
\bea
&&\hspace{-8mm}M_4^{(2)}-\frac{1}{2}\left(M_4^{(1)}\right)^2
\\ && \hspace{-8mm}
= \left( \kappa \over 8 \pi \right)^4 
\Bigg\{
us\Bigg[8(S_{1,3}(\frac{-s}{|t|})+\zeta_4)+\frac{\log^4(\frac{-s}{|t|})}{3}
+4\pi^2\left({\rm Li}_2(\frac{-s}{|t|})-\zeta_2+\frac{\log^2(\frac{-s}{|t|})}{2}
\right)\Bigg]
\nonumber \\ &&\hspace{-8mm}
+
tu\Bigg[8\left(S_{1,3}(-\frac{u}{s})+4\zeta_4-S_{1,3}(-1)-{\rm Li}_4(-\frac{u}{s})+\log(\frac{u}{s}){\rm Li}_3
(-\frac{u}{s})-\frac{1}{2}\log^2(\frac{u}{s}){\rm Li_2}(-\frac{u}{s})\right)
\nonumber\\ &&\hspace{-8mm}
+\frac{1}{3}\left(\log^4(\frac{|t|}{-s})+\log^4(\frac{u}{s})-4\log^3(\frac{u}{s})\log(\frac{|t|}{-s})\right)
+2\pi^2\log^2(\frac{|t|}{-u})+\frac{\pi^4}{3}\Bigg]
\Bigg\}
+ (u \leftrightarrow s) \,.
\nonumber
\label{reseucl}
\eea

\section{Infrared behavior and generalizations}
\setcounter{equation}{0}
\label{secIR}

In the previous section, we calculated the two-loop four-point
function in $\cN=8$ supergravity, 
and noted the particularly simple structure of its 
infrared divergences in eq.~(\ref{simple}).
In this section, we will derive 
the form of the leading-power divergence more heuristically,
in a way that can be generalized 
to higher $n$-point functions.
Before discussing ${\cN}=8$ supergravity, 
we briefly review IR divergences for ${\cN}=4$ SYM theories \cite{Bern:2005iz}
(for a review, see ref.~\cite{Dixon:2008tu}).

When we dimensionally regularize a theory in $D=4 - 2 \epsilon$ dimensions,
both UV and IR divergences appear as poles in $\epsilon$.
In a UV finite theory, such as $\cN=4$ SYM, the
poles in $\epsilon$ are solely due to IR divergences.
In gluon-gluon scattering in ${\cN}=4$ SYM, 
IR divergences arise both from soft gluons 
and from collinear gluons (which can exchange a virtual gluon
with soft transverse momentum), 
each of which gives rise to a $1/\epsilon$ pole at 1-loop, 
leading to a $1/\epsilon^2$ pole at that order. 
At $L$ loops, the leading IR divergence is therefore $\cO(1/\epsilon^{2L})$, 
arising from multiple soft gluon exchanges. 
In the large-$N$ (planar) limit, 
these IR divergences can be characterized by the Sudakov factor
$\cAdivSYM (s)$,
with one such factor for each pair of 
{\it adjacent} (external) gluons in the 
$n$-gluon amplitude,
\be
\prod_{i=1}^n \cAdivSYM (s_{i,i+1})
\ee
where $s_{i,i+1}=(k_i+k_{i+1})^2$. 
For $n=4$, this becomes $\cAdivSYM^2(s) \cAdivSYM^2(t)$,
where $s= s_{1,2} = s_{3,4}$ and $t = s_{2,3} = s_{4,1}$.
The Sudakov factor in the one-loop approximation is
\be
\label{sudakov}
\cAdivSYM (s) =\exp \left[
-\frac{\lambda_{SYM}} {(4\pi \epsilon)^2}
\left(\frac{\mu^2}{ -s}\right)^\epsilon+\cO(\lambda_{SYM}^2)\right]
\ee
where the SYM coupling $\lambda_{SYM}$ is 
the dimensionless 't Hooft coupling $g^2 N$.
The exponential Sudakov factor 
is modified at higher-loop order, 
but can be completely characterized by two functions 
of $\lambda_{SYM}$:
the cusp anomalous dimension $f(\lambda_{SYM})$ 
and the collinear anomalous dimension $g(\lambda_{SYM})$. 

Now consider ${\cN}=8$ supergravity, 
which is UV finite at least to third (maybe eighth) order 
in perturbation theory, and possibly to all 
orders \cite{Bern:2006kd,Green:2006gt,Green:2006yu,Bern:2007hh}.
Therefore, to at least third (maybe eighth) order,
the poles in the $\cN=8$ scattering amplitudes are
due only to IR divergences.
It has been known for some time that 
gravity theories have infrared divergences due to soft gravitons, 
but that collinear divergences are absent \cite{Weinberg:1965nx,Donoghue:1999qh}.
Hence at $L$ loops, IR divergences are expected 
to give rise to a leading $1/\epsilon^{L}$ divergence. 
This is borne out at two loops by the calculations of the last section,  
where the $1/\epsilon^4$ and $1/\epsilon^3$ poles cancel 
out of the final result (\ref{simple}).  
In ref.~\cite{Dunbar:1995ed}, Dunbar and Norridge showed that the 
one-loop amplitude has a $1/\epsilon$ divergence.

A deep relationship exists between the perturbative amplitudes 
of $\cN=8$ supergravity and $\cN=4$ SYM theory, 
going back to the work of Kawai, Lewellen, and Tye \cite{Kawai:1985xq}, 
and reviewed in ref.~\cite{Bern:1998ug}.
The one-loop amplitudes of $\cN=8$ supergravity and $\cN=4$ SYM
are expressed in terms of the same scalar integral $\cI_4^{(1)}(s,t)$,
and the IR divergences
are described by the same product of Sudakov factors at one loop,
with two differences. 
The first difference 
is that in gravity theories, there is no large-$N$ limit, 
so we must consider planar and non-planar graphs on the same footing. 
As a result, there is a factor of 
$\cAdiv(s)$
for {\it every} pair of external gravitons, 
not just adjacent gravitons
\be
\prod_{i<j} \cAdiv(s_{i,j}) \,.
\ee
For the four-point function, this becomes 
$\cAdiv^2(s) \cAdiv^2(t)\cAdiv^2(u)$
where 
$s= s_{1,2} = s_{3,4}$, 
$t = s_{2,3} = s_{1,4}$, and
$u = s_{1,3} = s_{2,4}$. 
The second difference is that the supergravity coupling 
\be
\lambda=\left( \kappa\over 2 \right)^2
\left( 4\pi \e^{-\gamma} \right)^\epsilon
\ee
is dimensionful, so the factor of $\lambda_{SYM}$ in the 
Sudakov factor (\ref{sudakov})
must be replaced  by the dimensionless effective coupling $\lambda \cdot s$. 
Hence, the IR divergent part of the four-graviton amplitude at one loop
is expected to be 
\bea
\label{gravsudakov}
&&\hspace{-15mm}
\cAdiv^2(s) \cAdiv^2(t)\cAdiv^2(u)
\nonumber\\&=&
\exp \left\{ 
\left[ -\frac{ 2 \lambda s}{(4 \pi \epsilon)^2}
\left(\frac{\mu^2}{-s}\right)^\epsilon
-\frac{ 2 \lambda t}{(4 \pi \epsilon)^2}
\left(\frac{\mu^2}{-t}\right)^\epsilon
-\frac{ 2 \lambda u}{(4 \pi \epsilon)^2}
\left(\frac{\mu^2}{-u}\right)^\epsilon \right] \Bigg|_{\rm divergent}
+\cO(\lambda^2)
\right\} 
\nonumber\\&=&
\exp \Bigg\{
\frac{\lambda}{8 \pi^2 \epsilon}
\left[ s\log\left(-s \over \mu^2\right) +t\log\left(-t\over \mu^2\right)
+u\log\left(-u\over \mu^2\right) \right] +\cO(\lambda^2) \Bigg\}
\eea
where the $1/\epsilon^2$ term vanishes because it is multiplied by $s+t+u=0$. 
Thus our heuristic argument reproduces the IR divergence of the 
one-loop amplitude (\ref{oneloopresulte}). 
The one-loop IR divergence (\ref{gravsudakov}) 
was obtained over a decade ago by Dunbar and Norridge \cite{Dunbar:1995ed}.

Analogously,
the IR divergent part of the $n$-graviton amplitude should
depend on the product of all distinct factors of $\cAdiv(s_{i,j})$, 
since again planar and nonplanar graphs are on equal footing and
since the same divergent function as for SYM appears in the scalar 
diagrams (due to the KLT relations). 
Therefore at one-loop, the IR divergent factor for the $n$-graviton 
amplitude is 
\bea
\label{ngravsudakov}
\prod_{i<j} \cAdiv(s_{i,j})
&=&
\exp \left\{ 
-\frac{\lambda }{(4 \pi \epsilon)^2}
\sum_{i<j} s_{i,j}
\left(\frac{\mu^2}{-s_{i,j}}\right)^\epsilon  \Bigg|_{\rm divergent}
+ \cO(\lambda^2) \right\} 
\nonumber\\&=&
\exp \Bigg\{ 
-\frac{\lambda}{8 \pi^2 \epsilon^2}
\sum_{i<j} k_i \cdot k_j
+\frac{\lambda }{16 \pi^2 \epsilon}
\sum_{i<j} s_{i,j}\log\left(\frac{-s_{i,j}}{\mu^2}\right)
+\cO(\lambda^2)
\Bigg\}
\nonumber\\ &=&
\exp \Bigg\{ 
 \frac{\lambda }{16 \pi^2 \epsilon}
\sum_{i<j} s_{i,j}\log\left(\frac{-s_{i,j}}{\mu^2}\right)
+\cO(\lambda^2)
\Bigg\}
\eea
where $s_{i,j}=(k_i + k_j)^2 = 2k_i\cdot k_j$ because external states 
are massless,
and the coefficient of $1/\epsilon^2$,
namely $\sum_{i<j} k_i \cdot k_j$,
vanishes for massless gravitons due to 
momentum conservation $\sum_{i=1}^n k_i=0$. 
The IR divergence of the one-loop $n$-graviton amplitude
was also obtained in ref.~\cite{Dunbar:1995ed}. 

The exponent of the SYM Sudakov factor (\ref{sudakov}) gets a
correction at $\cO(\lambda_{SYM}^2)$ due to the cusp anomalous dimension
$f(\lambda_{SYM})$.  
In principle, the analogous factors $\cAdiv(s)$ in eqs.~(\ref{gravsudakov}) 
and (\ref{ngravsudakov}) could get an $\cO(\lambda^2)$ IR divergent correction, 
but the two-loop calculation of the previous
section revealed the absence of such a correction. 
Because there is no analog of the function $f(\lambda_{SYM})$ 
for supergravity, $\cAdiv(s)$ differs from $\cAdivSYM (s)$ at higher orders.

The calculation of the previous section showed 
that the {leading} $1/\epsilon^2$ pole
of the two-loop four-point amplitude 
is indeed correctly given by eq.~(\ref{gravsudakov}),
with no $\cO(\lambda^2)$ modification.
One could reasonably conjecture that the {leading} $1/\epsilon^{2L}$ 
divergence for the $L$-loop four-point amplitude 
is also given by  eq.~(\ref{gravsudakov}),
namely
\be
{1 \over L!} \left( \lambda \over 8 \pi^2 \epsilon\right)^L 
\left[
 s\log\left(-s \over \mu^2\right)
+t\log\left(-t\over \mu^2\right)
+u\log\left(-u\over \mu^2\right)
\right]^L
+ \cO(1 / \epsilon^{L-1} )
\ee
and similarly that the
leading divergence for the $L$-loop $n$-point function 
is given by\footnote{As we can see from ref.~\cite{Dunbar:1995ed},
there is no fundamental difference between $n=4$ and $n>4$ amplitudes 
as far as IR divergences are concerned, 
so we can extend the $n>4$ results to the same loop order 
as the $n=4$ result.}
\be
{1 \over L!} \left( \lambda \over 16 \pi^2 \epsilon\right)^L 
\left[\sum_{i<j} s_{i,j}\log\left(\frac{-s_{i,j}}{\mu^2}\right)
\right]^L 
+ \cO(1 / \epsilon^{L-1} )
\ee
These are consistent with general expectations for the order 
of the leading divergence,
but we have not attempted to verify them beyond two loops.

We found in fact a stronger result for the four-point function at two loops;
namely, that {\it both} the leading $1/\epsilon^2$ 
{\it and} the subleading $1/\epsilon $ IR divergence
are given by the exponential of the one-loop amplitude (\ref{oneloopresult})
\be
\label{IRexp}
\cAdiv^2(s) \cAdiv^2(t)\cAdiv^2(u)
= \exp \left[ M_4^{(1)}(\epsilon) 
+\cO(\lambda^3) \right]  \Bigg|_{\rm divergent} \,.
\ee
This implies that the total two-loop divergence 
involves the finite as well as the divergent part of the exponent.
Equation (\ref{IRexp}) differs from $\cN=4$ SYM theory,  
in which the two-loop divergences are given by 
\be
\exp\left[  
a  M_4^{(1)}(\epsilon) - 
a^2 (\zeta_2 + \epsilon \zeta_3) M_4^{(1)} (2 \epsilon) 
+ \cO(a^3) \right] \Bigg|_{\rm divergent} \,,
\qquad
a =\left(\lambda_{SYM}\over 8\pi^2 \right) 
\left( 4\pi \e^{-\gamma} \right)^\epsilon
\ee
where the second term,
which contributes to the $1/\epsilon^2$ and $1/\epsilon$ divergences,
comes from the $\cO(\lambda_{SYM}^2)$ coefficients
of the anomalous dimensions $f(\lambda_{SYM})$ and $g(\lambda_{SYM})$.
An argument for the absence of an $\cO(\lambda^2)$ 
correction to the exponent in eq.~(\ref{IRexp}) would go as follows:  
due to the dimensionality of the coupling $\lambda$, 
the second term would have to be multiplied not 
by $\lambda$, but by some function of $\lambda s$, 
$\lambda t$, and $\lambda u$, but the only symmetric term
at  first order, $\lambda (s+t+u)$ vanishes.
At three loops, of course, a nonvanishing 
term $\lambda^2 (s^2 + t^2 + u^2)$ could in principle come in. 
The only allowed possibility that would 
mimic the ${\cN}=4$ SYM result, an 
$f(\lambda^2(s^2+t^2+u^2))M^{(1)}$ term, implies a factorization of the momentum dependence which seems 
unlikely. 
One cannot exclude,  however, the possibility that 
higher-order corrections do not organize into 
a single function, but give, \eg,  an infinite series 
$\sum_{n\geq 2}c_n\lambda^n(s^n+t^n+u^n)M^{(1)}$.  After all, ${\cN}=8$ supergravity is potentially nonrenormalizable, being a field theory of quantum gravity.

On the more optimistic side,  
it remains possible that the simple behavior 
in eq.~(\ref{IRexp}) continues at higher loops
(at least up to the order to which the theory is UV finite),
and that the IR divergences (both leading and subleading)
of the four-point function are exactly given by 
\be
\cAdiv^2(s) \cAdiv^2(t)\cAdiv^2(u)
= \exp \left[ M_4^{(1)}(\epsilon) \right]  
\Bigg|_{\rm divergent}
\label{fourpointdiv}
\ee
to all orders in the coupling $\lambda$.
An even more daring conjecture is that the complete IR divergences of 
the $n$-point amplitudes
are given by the exponential of the 1-loop amplitude
\be
\prod_{i<j} \cAdiv(s_{i,j}) = \exp \left[ M_n^{(1)}(\epsilon) \right]
\Bigg|_{\rm divergent}
\label{npointdiv}
\ee
to all orders in the coupling $\lambda$.
In principle, the expressions for the IR-divergent 
contributions (\ref{fourpointdiv}) and (\ref{npointdiv})
could be modified\footnote{We
thank both Lance Dixon and the referee for pointing this out.}
by functions $E_n(\epsilon)$ that vanish as 
$\epsilon \to 0$,
as in the case of $\cN=4$ SYM theory \cite{Bern:2005iz}.
On the other hand, such functions $E_n(\epsilon)$ 
could be absorbed into the $\epsilon$-expansions of $f^{(l)}(\epsilon)$ 
\cite{Bern:2005iz}
and thus related to the anomalous dimensions.
As we pointed out in the previous paragraph, 
such ``anomalous dimension-like" terms
may well be absent in supergravity.

In summary, we have found similarities between the 
IR divergences of ${\cN}=8$ supergravity and those of 
planar ${\cN}=4$ SYM, as well as significant differences, due to the absence of collinear divergences, and 
due to the presence of a dimensionful coupling constant
for supergravity.

\section{Conclusions}
\setcounter{equation}{0}
\label{secconcl}

In this paper, the one- and two-loop graviton four-point amplitudes 
in ${\cN}=8$ supergravity were explicitly computed.
A number of regularities appeared, 
most importantly an ABDK-like relation (\ref{finalresult})
between the one- and two-loop amplitudes.

Specifically, we found that the IR divergent part of the 
two-loop amplitude is the divergent part of one-half the square 
of the {full} one-loop amplitude, 
suggesting an exponentiation of the IR divergences. 
We gave a heuristic argument for the IR divergences
of graviton scattering amplitudes which allows a generalization 
to one-loop $n$-point amplitudes, 
and a conjectured generalization to $L$-loop $n$-point amplitudes.

Moreover, most of the finite part of the two-loop amplitude 
also comes from the square of the full one-loop amplitude
(i.e., including the order $\epsilon$ part), 
with a very simple remainder. 
This is reminiscent of ${\cN}=4$ SYM,
where the presence of the dual conformal symmetry of the dual Wilson loop 
restricts the form of the $n=4$ and $n=5$ 
amplitudes \cite{Drummond:2007cf,Drummond:2007au} 
to the BDS exponential form 
(essentially the exponential of the one-loop amplitude, 
together with the extra information contained in the functions
$f(\lambda_{SYM})$ and $g(\lambda_{SYM})$).
But for the $n=6$ amplitude, dual conformal symmetry does not 
fix the result, 
and it was found that at two-loops, 
besides the BDS exponential form, 
there is a small remainder 
function \cite{Drummond:2007bm,Bern:2008ap,Drummond:2008aq} 
(small means, \eg, that it does not affect Regge 
behavior \cite{Brower:2008nm} \footnote{
But see ref.~\cite{Bartels:2008ce}.}
and it arises as a correction \cite{Itoyama:2008je}). 
One could expect that something similar is at work here
(for supergravity there is no dual conformal symmetry 
to fix the amplitude): 
since there are no analogs of $f(\lambda_{SYM})$ and $g(\lambda_{SYM})$ 
due to the dimensionality of the coupling, 
the amplitude is given by the exponential of the one-loop amplitude, 
with a simple finite remainder
(at least to the order to which ${\cN}=8$ supergravity is finite).

The discussion in this paper does not assume that 
${\cN}=8$ supergravity is or is not perturbatively UV finite. 
If UV finiteness breaks down at $L$-loops, 
then our conjectures could nonetheless be valid up to that loop level.

\vskip 5mm

{\bf Acknowledgments} 

HN would like to thank Radu Roiban and George Sterman for discussions. 
HJS thanks Lance Dixon for conversations, 
and for informing us of his own work on the subject. 
All of the authors wish to thank Lance Dixon for a careful
reading of and comments on our manuscript,
and for confirming that eq.~(\ref{finalresult}) 
is compatible with results he obtained independently.
HN's research  has been done with partial support 
from  MEXT's program ``Promotion of Environmental Improvement for 
Independence of Young Researchers" under the Special Coordination 
Funds for Promoting Science and Technology. 

\section*{Appendix}
\setcounter{equation}{0}
\def\theequation{A.\arabic{equation}}

The generalized polylogarithms of Nielsen are defined by \cite{Kolbig} 
\be
\label{nielsen}
S_{n,p}(x)  = \frac{(-1)^{n+p-1}}{(n-1)!\, p!} \int_0^1 dt \,
  \frac{\log^{n-1}(t) \log^{p}(1-xt)}{t}, \quad \quad \quad n,p \ge 1, \ \
  x\le 1
\ee
which in the case of $p=1$ reduce to the usual polylogarithms 
\be
S_{n-1,1} (x)  \equiv  \Li{n}{x} 
\ee
For $n=1$, eq.~(\ref{nielsen}) may be rewritten as
\be
S_{1,p}(x)=\int_0^x\frac{dz}{z}\frac{(-\log(1-z))^p}{p!}
\,.
\label{polyl}
\ee
The following identities for generalized polylogarithms are valid
for $x<0$, with $y = 1/x$ and $L = - \log(-x) = \log(-y)$:
\bea
\label{appone}
\Li{2}{x} &=&  - \Li{2}{y} -\frac{1}{2}L^2 -\frac{\pi ^2}{6}
\nonumber\\
\Li{3}{x} &=&    \Li{3}{y} + \frac{1}{6}L^3 +\frac{\pi ^2}{6} L
\nonumber\\
S_{1,2}(x) &=&  - S_{1,2}(y)  + \Li{3}{y} - L  \,\Li{2}{y}
-\frac{1}{6} L^3 +\zeta_3
\\
\Li{4}{x} &=&  - \Li{4}{y}
 -\frac{1}{24} L^4 -\frac{\pi ^2}{12} L^2 -\frac{7 \pi ^4}{360}
\nonumber\\
S_{2,2}(x) &=&   S_{2,2}(y) -2 \, \Li{4}{y} + L\, \Li{3}{y} 
+ \frac{1}{24} L^4 -\zeta_3 L  -\frac{7 \pi ^4}{360}
\nonumber\\
S_{1,3}(x) &=&   -S_{1,3}(y) 
+S_{2,2}(y)
-\Li{4}{y}
-L\,S_{1,2}(y) 
+L\,\Li{3}{y}
-\frac{1}{2} L^2 \, \Li{2}{y} 
-\frac{1}{24}L^4
-\frac{\pi ^4}{90}
\nonumber
\eea
and are used in sec.~2 to convert
eq.~(\ref{twoboxu}) to (\ref{twoboxt})  and 
eq.~(\ref{oneboxu}) to (\ref{oneboxt}).
The polylogarithms also obey the following identity 
when $0<y<1$  
\be
S_{n,p}(1-y) = \sum_{s=0}^{n-1} \frac{\log^s(1-y)}{s!}
\left[
S_{n-s,p}(1) 
-\sum_{r=0}^{p-1} \frac{(-\log y)^r}{r!}\,S_{p-r,n-s}(y) \right]
+  \frac{(-1)^p}{n!\, p!} \log^n(1-y)\log^p y
\ee
which becomes,
where $T = -\log y$ and $V = \log(1-y)$,
\bea
\label{apptwo}
\Li{2}{1-y} &=&  - \Li{2}{y}  + V T + \frac{\pi ^2}{6}
\nonumber\\
\Li{3}{1-y} &=&  - S_{1,2}(y)  - V\, \Li{2}{y} + \frac{1}{2} T V^2 
+\frac{\pi^2}{6} V  +\zeta_3
\nonumber\\
S_{1,2}(1-y)&=&   -\Li{3}{y} - T \, \Li{2}{y} + \frac{1}{2} V T^2 + \zeta_3
\\
\Li{4}{1-y} &=&  -  S_{1,3}(y) - V\, S_{1,2}(y)  
- \frac{1}{2} V^2 \, \Li{2}{y}
+ \frac{1}{6} T V^3
+ \frac{\pi^2}{12} V^2 
+ \zeta_3 V 
+\frac{\pi ^4}{90}
\nonumber\\
S_{2,2}(1-y) &=&   - S_{2,2}(y) -T\, S_{1,2}(y) - V\, \Li{3}{y} 
- V T\, \Li{2}{y} + \frac{1}{4} V^2 T^2  +\zeta_3 V   +\frac{\pi ^4}{360}
\nonumber\\
S_{1,3}(1-y) &=&  - \Li{4}{y} - T\, \Li{3}{y} - \frac{1}{2} T^2 \, \Li{2}{y}
 +\frac{1}{6} V T^3 + \frac{\pi ^4}{90}
\nonumber
\eea
which  are used in sec.~2 to obtain eqs.~(\ref{simple}) 
and (\ref{oneloopdiff}).

In eq.~(\ref{ftsu}), we obtained the expression 
\be 
\label{ftsuagain}
\f(t,s,u) =
8 \, S_{1,3} (y) + \frac{1}{3} \log^4 y + \frac{4 \pi^4}{45} 
+ i \pi \left[-8\, S_{1,2} (y) + \frac{4}{3} \log^3 y + 8 \zeta_3 \right],
\qquad y = - \frac{s}{t}
~~
\ee
valid in the region $t>0$ and  $s$, $u<0$,    
for the function that appears in eq.~(\ref{difference}),
the difference between the two-loop amplitude 
and one-half the square of the one-loop amplitude.
To obtain the full result for the difference,
we need to analytically continue $\f(t,s,u)$ to other regions.

To analytically continue $\f(t,s,u)$ to the region $s>0$ and $t$, $u<0$,
we let $s$ and $t$ traverse the upper half plane (in opposite directions), 
which causes $y = -s/t$ to go from a point between 0 and 1 on the real axis 
clockwise through an angle $2 \pi$ around the origin,
ending up at a point to the right of 1 on the real axis.
Hence,  
$\log y \to  \log y - 2 \pi i$ and $S_{n,p}(y) \to S_{n,p}(y + i 0)$.
Next, we use eqs.~(A.12) and (A.13) from  ref.~\cite{Anastasiou:1999bn}
to re-express this as
\bea
\f(t,s,u) &=&
-8 \,\Li{4}{x} +8 \,S_{2,2}(x) -8 \,S_{1,3}(x)
+8 \log x \, \Bigl( \Li{3}{x} - S_{1,2}(x)  \Bigr) -2 \pi ^2 \log^2 x 
\nonumber\\&& 
-4 \Bigl( \log^2 x  + \pi ^2 \Bigr) \Li{2}{x} -\frac{13 \pi ^4}{3}
+i \pi \left[  \frac{4}{3} \log^3 x+ \frac{8 \pi^2}{3} \log x  \right],
\qquad x= -\frac{t}{s}
\hspace{10mm}
\eea
valid for $s>0$ and $t$, $u<0$ (that is, for  $0<x<1$).
Finally, we simply let $x \to y$ to obtain 
\bea
\label{fstu}
\f(s,t,u) &=&
-8 \,\Li{4}{y} +8 \,S_{2,2}(y) -8 \,S_{1,3}(y)
+8 \log y \, \Bigl( \Li{3}{y} - S_{1,2}(y)  \Bigr) -2 \pi ^2 \log^2 y 
\nonumber\\&& 
-4 \Bigl( \log^2 y  + \pi ^2 \Bigr) \Li{2}{y} -\frac{13 \pi ^4}{3}
+i \pi \left[  \frac{4}{3} \log^3 y + \frac{8 \pi^2}{3} \log y  \right],
\qquad y= -\frac{s}{t}
\hspace{10mm}
\eea
valid for the region $t>0$ and  $s$, $u<0$.

To analytically continue $\f(t,s,u)$ to the region $u>0$ and  $s$, $t<0$,
we let $u$ and $t$ traverse the upper half plane (in opposite directions), 
which causes $y = -s/t$ to go from a point between 0 and 1 on the real axis 
clockwise through an angle $2 \pi$ around the point $y=1$,
ending up at a point on the negative real axis.
As a result 
\bea
\log y &\to&  \log (-y) -   i \pi 
\nonumber\\
S_{1,2}(y) &\to& 
S_{1,2}(y) -2 \pi^2 \log (-y)
+i \pi \left[ 2\, \Li{2}{y} +\frac{5 \pi^2}{3} \right]
\nonumber\\
S_{1,3}(y) &\to& 
S_{1,3}(y) -2 \pi^2 \, \Li{2}{y} -\pi ^4
+ i \pi \left[ 2\, S_{1,2}(y) -\frac{4\pi^2}{3} \log (-y) -2 \zeta_3
\right]\,.
\eea
Inserting these into eq.~(\ref{ftsuagain}), we obtain
\be
\f(t,s,u) =
8 \, S_{1,3}(y)
+ \frac{1}{3} \log^4(-y)
+2 \pi^2 \log^2(-y)
+\frac{199 \pi ^4}{45}
+i \pi \left[ 8 S_{1,2}(y) 
+\frac{8}{3} \pi^2 \log (-y)
-8 \zeta_3 \right]
\ee
valid for $u>0$ and $s$, $t<0$ (that is, for  $y<0$).
Then, to obtain $\f(s,u,t)$ for $t>0$ and $s$, $u<0$, 
we permute $s \to u \to t \to s$,
which takes $y \to (y-1)/y$.
Polylogarithms with argument $(y-1)/y$ can be expressed as 
polylogarithms with argument  $y/(y-1)$ by using eqs.~(\ref{appone}),
and the latter can be expressed as 
polylogarithms with argument $y$ by using eqs.~(A.15) through (A.20)
of ref.~\cite{Anastasiou:1999bn}, resulting in 
\bea
\label{fsut}
\f(s,u,t) &=&
8 \, \Li{4}{y} -8 \log y \, \Li{3}{y} +4 \log^2 y \, \Li{2}{y}  
+ \frac{1}{3} \log^4 (1-y)
-\frac{4}{3} \log y \log^3 (1-y)
\nonumber \\
&+&2 \log ^2 y  \log^2 (1-y)
+2 \pi^2 \Bigl[ \log (1-y) - \log y \Bigr]^2
+\frac{13 \pi ^4}{3}
+i \pi \Big[ -8 \, \Li{3}{y} 
\nonumber \\
&+&
  8 \log y \, \Li{2}{y}
+  4  \log^2 y \log (1-y)
-\frac{4}{3} \log^3 y
+\frac{8\pi^2}{3}  \log (1-y)
-\frac{8\pi^2}{3}  \log y
\Big]
\nonumber\\
\eea
valid for $t>0$ and $s$, $u<0$. 
Finally, we add eqs.~(\ref{fstu}) and (\ref{fsut})
to obtain the coefficient of $tu$ in eq.~(\ref{finalresult})


\begin{thebibliography}{10}

\bibitem{Anastasiou:2003kj}
C.~Anastasiou, Z.~Bern, L.~J. Dixon, and D.~A. Kosower, ``{Planar amplitudes in
  maximally supersymmetric Yang-Mills theory}.'' Phys. Rev. Lett. {\bf 91}
  (2003) 251602, \href{http://xxx.lanl.gov/abs/hep-th/0309040}
{{\tt hep-th/0309040}}. 

\bibitem{Bern:2005iz}
Z.~Bern, L.~J. Dixon, and V.~A. Smirnov, ``{Iteration of planar amplitudes in
  maximally supersymmetric Yang-Mills theory at three loops and beyond}.''
  Phys. Rev. {\bf D72} (2005) 085001,
  \href{http://xxx.lanl.gov/abs/hep-th/0505205}{{\tt hep-th/0505205}}.

\bibitem{Magnea:1990zb}
L.~Magnea and G.~Sterman, ``Analytic continuation of the Sudakov form-factor in
  QCD.'' Phys. Rev. {\bf D42} (1990) 4222--4227. 

\bibitem{Sterman:2002qn}
G.~Sterman and M.~E. Tejeda-Yeomans, ``Multi-loop amplitudes and resummation.''
  Phys. Lett. {\bf B552} (2003) 48--56,
  \href{http://xxx.lanl.gov/abs/hep-ph/0210130}{{\tt hep-ph/0210130}}.

\bibitem{Alday:2007hr}
L.~F. Alday and J.~Maldacena, ``Gluon scattering amplitudes at strong
  coupling.'' JHEP {\bf 06} (2007) 064,
  \href{http://xxx.lanl.gov/abs/arXiv:0705.0303 [hep-th]}{{\tt arXiv:0705.0303
  [hep-th]}}. 

\bibitem{Beisert:2006ez}
N.~Beisert, B.~Eden, and M.~Staudacher, ``{Transcendentality and crossing}.''
  J. Stat. Mech. {\bf 0701} (2007) P021,
  \href{http://xxx.lanl.gov/abs/hep-th/0610251}{{\tt hep-th/0610251}}.

\bibitem{Drummond:2007cf}
J.~M. Drummond, J.~Henn, G.~P. Korchemsky, and E.~Sokatchev, ``{On planar gluon
  amplitudes/Wilson loops duality}.''
  \href{http://xxx.lanl.gov/abs/arXiv:0709.2368 [hep-th]}{{\tt arXiv:0709.2368
  [hep-th]}}. 

\bibitem{Drummond:2007au}
J.~M. Drummond, J.~Henn, G.~P. Korchemsky, and E.~Sokatchev, ``{Conformal Ward
  identities for Wilson loops and a test of the duality with gluon
  amplitudes}.'' \href{http://xxx.lanl.gov/abs/arXiv:0712.1223 [hep-th]}
{{\tt arXiv:0712.1223 [hep-th]}}. 

\bibitem{Alday:2007he}
L.~F. Alday and J.~Maldacena, ``{Comments on gluon scattering amplitudes via
  AdS/CFT}.'' JHEP {\bf 11} (2007) 068,
  \href{http://xxx.lanl.gov/abs/0710.1060}{{\tt arXiv:0710.1060 [hep-th]}}. 



\bibitem{Drummond:2007bm}
J.~M. Drummond, J.~Henn, G.~P. Korchemsky, and E.~Sokatchev, ``{The hexagon
  Wilson loop and the BDS ansatz for the six- gluon amplitude}.''
  \href{http://xxx.lanl.gov/abs/arXiv:0712.4138 [hep-th]}{{\tt arXiv:0712.4138
  [hep-th]}}. 

\bibitem{Bern:2008ap}
Z. Bern, L. J. Dixon, D. A. Kosower, R. Roiban, M. Spradlin, 
C. Vergu, A. Volovich, 
``{The Two-Loop Six-Gluon MHV Amplitude in Maximally
  Supersymmetric Yang-Mills Theory}.''
  \href{http://xxx.lanl.gov/abs/0803.1465}{{\tt arXiv:0803.1465 [hep-th]}}. 

\bibitem{Drummond:2008aq}
J.~M. Drummond, J.~Henn, G.~P. Korchemsky, and E.~Sokatchev, ``{Hexagon Wilson
  loop = six-gluon MHV amplitude}.''
  \href{http://xxx.lanl.gov/abs/0803.1466}{{\tt arXiv:0803.1466 [hep-th]}}. 

\bibitem{Astefanesei:2007bk}
D.~Astefanesei, S.~Dobashi, K.~Ito, and H.~S. Nastase, ``{Comments on gluon
  6-point scattering amplitudes in N=4 SYM at strong coupling}.'' JHEP {\bf 12}
  (2007) 077, \href{http://xxx.lanl.gov/abs/arXiv:0710.1684 [hep-th]}
{{\tt arXiv:0710.1684 [hep-th]}}. 

\bibitem{Buchbinder:2007hm}
E.~I. Buchbinder, ``{Infrared Limit of Gluon Amplitudes at Strong Coupling}.''
  Phys. Lett. {\bf B654} (2007) 46--50,
  \href{http://xxx.lanl.gov/abs/0706.2015}{{\tt arXiv:0706.2015 [hep-th]}}. 

\bibitem{Bern:2007hh}
Z.~Bern {\em et.~al.}, ``{Three-Loop Superfiniteness of N=8 Supergravity}.''
  Phys. Rev. Lett. {\bf 98} (2007) 161303,
  \href{http://xxx.lanl.gov/abs/hep-th/0702112}{{\tt hep-th/0702112}}.

\bibitem{Green:2006yu}
M.~B. Green, J.~G. Russo, and P.~Vanhove, ``{Ultraviolet properties of maximal
  supergravity}.'' Phys. Rev. Lett. {\bf 98} (2007) 131602,
  \href{http://xxx.lanl.gov/abs/hep-th/0611273}{{\tt hep-th/0611273}}.

\bibitem{Bern:2006kd}
Z.~Bern, L.~J. Dixon, and R.~Roiban, ``{Is N = 8 supergravity ultraviolet
  finite?}.'' Phys. Lett. {\bf B644} (2007) 265--271,
  \href{http://xxx.lanl.gov/abs/hep-th/0611086}{{\tt hep-th/0611086}}.

\bibitem{Green:2006gt}
M.~B. Green, J.~G. Russo, and P.~Vanhove, ``{Non-renormalisation conditions in
  type II string theory and maximal supergravity}.'' JHEP {\bf 02} (2007) 099,
  \href{http://xxx.lanl.gov/abs/hep-th/0610299}{{\tt hep-th/0610299}}.

\bibitem{Green:2007zzb}
M.~B. Green, H.~Ooguri, and J.~H. Schwarz, ``{Decoupling Supergravity from the
  Superstring}.'' Phys. Rev. Lett. {\bf 99} (2007) 041601,
  \href{http://xxx.lanl.gov/abs/0704.0777}{{\tt arXiv:0704.0777 [hep-th]}}. 

\bibitem{Kawai:1985xq}
H.~Kawai, D.~C. Lewellen, and S.~H.~H. Tye, ``{A Relation Between Tree
  Amplitudes of Closed and Open Strings}.'' Nucl. Phys. {\bf B269} (1986) 1.

\bibitem{Bern:1998ug}
Z.~Bern, L.~J. Dixon, D.~C. Dunbar, M.~Perelstein, and J.~S. Rozowsky, ``{On
  the relationship between Yang-Mills theory and gravity and its implication
  for ultraviolet divergences}.'' Nucl. Phys. {\bf B530} (1998) 401--456,
  \href{http://xxx.lanl.gov/abs/hep-th/9802162}{{\tt hep-th/9802162}}.

\bibitem{Tausk:1999vh}
J.~B. Tausk, ``{Non-planar massless two-loop Feynman diagrams with four
  on-shell legs}.'' Phys. Lett. {\bf B469} (1999) 225--234,
  \href{http://xxx.lanl.gov/abs/hep-ph/9909506}{{\tt hep-ph/9909506}}.

\bibitem{Smirnov:1999gc}
V.~A. Smirnov, ``{Analytical result for dimensionally regularized massless
  on-shell double box}.'' Phys. Lett. {\bf B460} (1999) 397--404,
  \href{http://xxx.lanl.gov/abs/hep-ph/9905323}{{\tt hep-ph/9905323}}.

\bibitem{Dixon:2008tu}
L.~J. Dixon, ``{Gluon scattering in N=4 super-Yang-Mills theory from weak to
  strong coupling}.'' \href{http://xxx.lanl.gov/abs/0803.2475}
{{\tt arXiv:0803.2475 [hep-th]}}. 

\bibitem{Weinberg:1965nx}
S.~Weinberg, ``{Infrared photons and gravitons}.'' Phys. Rev. {\bf 140} (1965)
  B516--B524. 

\bibitem{Donoghue:1999qh}
J.~F. Donoghue and T.~Torma, ``{Infrared behavior of graviton-graviton
  scattering}.'' Phys. Rev. {\bf D60} (1999) 024003,
  \href{http://xxx.lanl.gov/abs/hep-th/9901156}{{\tt hep-th/9901156}}.

\bibitem{Dunbar:1995ed}
D.~C. Dunbar and P.~S. Norridge, ``{Infinities within graviton scattering
  amplitudes}.'' Class. Quant. Grav. {\bf 14} (1997) 351--365,
  \href{http://xxx.lanl.gov/abs/hep-th/9512084}{{\tt hep-th/9512084}}.

\bibitem{Brower:2008nm}
R.~C. Brower, H.~Nastase, H.~J. Schnitzer, and C.-I. Tan, ``{Implications of
  multi-Regge limits for the Bern-Dixon- Smirnov conjecture}.''
  \href{http://xxx.lanl.gov/abs/0801.3891}{{\tt arXiv:0801.3891 [hep-th]}}. 

\bibitem{Bartels:2008ce}
J.~Bartels, L.~N. Lipatov, and A.~S. Vera, ``{BFKL Pomeron, Reggeized gluons
  and Bern-Dixon-Smirnov amplitudes}.''
  \href{http://xxx.lanl.gov/abs/0802.2065}{{\tt arXiv:0802.2065 [hep-th]}}. 

\bibitem{Itoyama:2008je}
H.~Itoyama, A.~Mironov, and A.~Morozov, ``{'Anomaly' in n=infinity
  Alday-Maldacena Duality for Wavy Circle}.''
  \href{http://xxx.lanl.gov/abs/0803.1547}{{\tt arXiv:0803.1547 [hep-th]}}.

\bibitem{Kolbig}
K.S.~K\"olbig, J.A.~Mignaco and E.~Remiddi,
{\it BIT Numerical Mathematics} {\bf 10} (1970) 38.

\bibitem{Anastasiou:1999bn}
C.~Anastasiou, E.~W.~N. Glover, and C.~Oleari, ``{The two-loop scalar and
  tensor pentabox graph with light- like legs}.'' Nucl. Phys. {\bf B575} (2000)
  416--436, \href{http://xxx.lanl.gov/abs/hep-ph/9912251}{{\tt hep-ph/9912251}}. 

\end{thebibliography}

\providecommand{\href}[2]{#2}\begingroup\raggedright\endgroup

\end{document}